\documentclass[12pt,titlepage]{article}
\usepackage{epsfig}
\usepackage{graphicx}
\usepackage{amsmath,amssymb}
\usepackage{bm}
\usepackage{color}
\usepackage{empheq}
\usepackage{slashed}
\usepackage{cite}
\usepackage[all]{xy}
\usepackage{comment}
\newcommand{\beq}{\begin{equation}}
\newcommand{\eeq}{\end{equation}}
\newcommand{\bea}{\begin{eqnarray}}
\newcommand{\eea}{\end{eqnarray}}
\newcommand{\bean}{\begin{eqnarray*}}
\newcommand{\eean}{\end{eqnarray*}}
\newcommand{\vX}{{\vec {\bm X}}}

\newcommand{\wD}{{\widetilde D}}
\newcommand{\wS}{{ {\bm S}}}
\newcommand{\ws}{{{S}}}

\newcommand{\bdelta}{{\bm \delta}}

\newcommand{\WJ}{J}
\newcommand{\WQ}{Q}
\newcommand{\NQ}{{\cal Q}}
\newcommand{\Nq}{\rho}

\title{
\vskip -150pt
\flushright{\small DIAS-STP-18-06}\\
\vskip 150pt
\begin{center}
On the definition of mass in general relativity: Noether charges and conserved quantities in diffeomorphism invariant theories
\end{center}
}

\author{{Brian P. Dolan}\\
{\small Department of Theoretical Physics, National University
of Ireland,}\\
{\small  Maynooth, Ireland}\\
{\small\textit {and}}\\
{\small Dublin Institute for Advanced Studies,
10 Burlington Rd., Dublin, Ireland}\\
{\small e-mail: \texttt{bdolan@thphys.nuim.ie}}}

\begin{document}

\maketitle


\begin{abstract}

Geometrically the phase space of a mechanical system involves the co-tangent bundle of the configuration space.
The phase space of a relativistic field theory is infinite dimensional and can be endowed with a symplectic structure defined in a perfectly co-variant manner that is very useful for discussing symmetries and conserved quantities of the system. In general relativity the symplectic structure takes Darboux form and it is shown in this work that the presence of a cosmological constant does not change this conclusion.
  
For space-times that admit time-like Killing vectors the formalism can be used to define mass in general relativity and it is known that, for asymptotically flat black holes, this mass is identical the the usual Arnowitt-Desner-Misner mass while for asymptotically anti-de Sitter Kerr metrics it is the same as the Henneaux-Teitelboim mass.  We show that the same formalism can also be used to derive the Brown-York mass and the Bondi mass for stationary space times, in particular the Brown-York mass has a natural interpretation in terms of differential forms on the space of solutions of the theory.

\end{abstract}
\section{Introduction}

Some years ago Crnkovi\'c and Witten  \cite{C-W} gave a method for constructing a symplectic form on the space of solutions, ${\cal S}$, of the equations of motion of a relativistic field theory.  They used their formalism to obtain the relevant symplectic forms for Yang-Mills theory and for general relativity.  Their construction provides a co-variant description of relativistic field theories in the phase space of solutions modulo gauge transformations (and diffeomorhpisms)
${\cal G}$, $\widehat {\cal S}={\cal S}/{\cal G}$,
which is ideally suited to studying symmetries and conserved quantities.

The idea of a symplectic structure for diffeomorphism invariant theories 
was first introduced in \cite{Friedman} to investigate
instabilities in rotating relativistic fluids. 
Wald and collaborators have generalized Crnkovi\'c and Witten's formalism to
a very wide class of diffeomorphism invariant theories in \cite{Lee+Wald,Wald1,Wald2, Iyer+Wald} and studied conserved quantities associated with Killing symmetries, such as angular momentum in rotationally invariant solutions and mass in stationary solutions.

We first summarise the construction of the symplectic form and the role of diffeomorphisms and Killing symmetries in general.  Examples of the statements made in the introduction are given in the main text following.
One starts with an $(n+1)$-dimensional space-time manifold ${\cal M}$
with boundary ${\partial M}$.
The space-time comes with a metric, and possibly other fields such as Yang-Mills fields, and the space of all field configurations ${\cal F}$ is infinite
dimensional.  The dynamics is determined by a
variational principle with a Lagrangian $L$, which is a gauge invariant
$(n+1)$-form on ${\cal M}$, and an action
\[
  {\cal A}[{\cal F}]=\int_{\cal M} L({\cal F})\]
which is a diffeomorphism invariant functional of the fields.
A solution of the equations of motion is a field configuration that
extremises ${\cal A}$.

It is assumed that space-time can be foliated using a time parameter $t$ and that
surfaces of constant $t$ are space-like hypersurfaces, $\Sigma_t$. 
An infinitesimal variation of any solution of the equations of motion
that satisfies the linearised equations of motion is
a 1-form on ${\cal S}$, more correctly a cross-section of the co-tangent bundle
$T^*{\cal S}$. 

A symplectic form on the space of solutions is obtained by using $L$
to construct an $1$-form\footnote{In this general discussion bold face symbols will represent forms on ${\cal S}$.} $\bm\theta$ on ${\cal S}$, which is also an $n$-form on ${\cal M}$, and is a pre-symplectic potential on $\Sigma_t$,
\textit{i.e.}$\;{\bm\theta}$ does not itself furnish
a symplectic potential on the space of solutions
(it is not necessarily diffeomorphism invariant) but it gives one when
diffeomorphisms are modded out.
When $\Sigma_t$ is a Cauchy surface,
\[{\bm \Theta}=\int_{\Sigma_t}\bm{{\bm \theta}}\]
gives a pre-symplectic potential on ${\cal S}$. 
Under a second independent variation of the dynamical fields one obtains another $n$-form on ${\cal M}$
\[\bm{\omega}={\bm \delta} \bm{{\bm \theta}}\]
where ${\bm \delta}$ is the exterior derivative on the space of solution ${\cal S}$.  
Since ${\bm \delta}^2=0$  
\[{\bm \delta} \bm{\omega}=0\] and
$\bm{\omega}$  is a pre-symplectic density in the sense that 
\[\bm{\Omega}= \int_{\Sigma_t}\bm{\omega}\] is a pre-symplectic $2$-form on ${\cal S}$.\footnote{${\bm \delta} \int_{\Sigma_t} = \int_{\Sigma_t} {\bm \delta}$ since,
while  $\Sigma_t$ depends on the co-ordinates, it is independent of the fields, in particular of the metric.}  It is called pre-symplectic because it is not a genuine symplectic density, it is not necessarily gauge and diffeomorphism invariant

There is a very elegant interplay between the $d$-cohomology on ${\cal M}$ and the ${\bm \delta}$-cohomology on ${\cal S}$.
The construction is such that $\bm{\omega}$ is not only closed as a $1$-form on ${\cal S}$
but also as an $n$-form on ${\cal M}$,  $d\bm{\omega}=0$, hence
\[ \int_{\cal M} d \bm\omega=\int_{\partial {\cal M}} \bm{\omega}=0.\]  
If ${\cal M}$ has the topology $T\times \Sigma$, where $T=[t,t']\subset R$
is a time interval and $\Sigma$ is a compact Cauchy hypersurface without boundary, then $\partial {\cal M}$ consists of 2 copies of $\Sigma$.
Then foliate ${\cal M}$, using $t$ as a time-parameter,
and $\int_{\Sigma_t}{\bm \omega}$ is independent of the value of $t$ chosen so we can drop the subscript $t$ and
\[ \bm{\Omega}=\int_{\Sigma}\bm{\omega}\] 
is independent of the Cauchy hypersurface $\sigma$.

\begin{figure}[h]\label{space-time-tube}
\centerline{\includegraphics[scale=0.7]{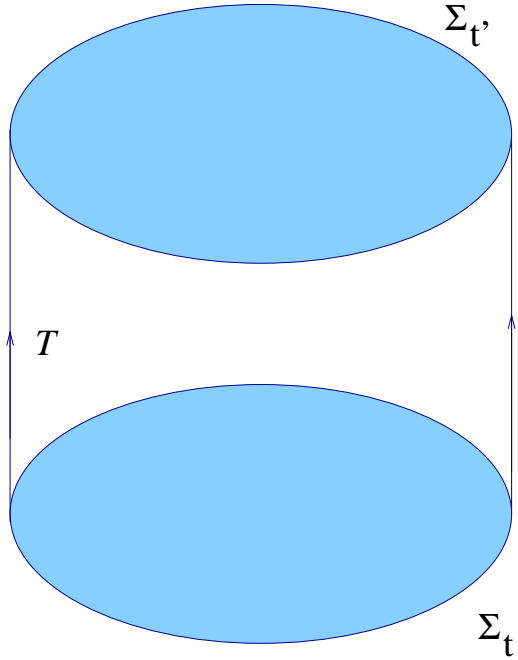}}
\caption{the boundary of ${\cal M}$ consists of two space-like hypersurfaces
  $\Sigma$ and $\Sigma'$ connected by a time-like tube ${\mathit T}$, with
  ${\it T}=\partial \Sigma \times [t,t']$.}
\end{figure}

Under a projection from the space of solutions ${\cal S}$ to 
the space of solutions modulo gauge transformations and diffeomorphisms
${\widehat {\cal S}}$
the symplectic form $\widehat {\bm \Omega}$ on  $\widehat{\cal S}$
must pull-back to the pre-symplectic form ${\bm \Omega}$ on ${\cal S}$ . 
This will be the case if
\begin{itemize}
  \item ${\bm \omega}$ is gauge invariant and $d$-exact whenever one of the metric variations is a diffeomorphism;
  \item  $\Sigma$ is compact without boundary.
    \end{itemize}
    Then it is shown in \cite{Wald2} that, when one of the metric variations corresponds to a diffeomorphism generated by a vector field $\vec X$,
the dependence of ${\bm \omega}$ on $\vec  X$ is such that ${\bm \omega}(\vec  X)$ is not only $d$-closed as an $n$-form on ${\cal M}$ but it is also
$d$-exact\footnote{The condition ${\bm \delta} {\bm \theta} =d {\bm \phi}$ is reminiscent of the Stora-Zumino descent equations
  in the study of anomalies \cite{Stora-Zumino}. Indeed the whole formalism
  is intimately related to a cohomological structure that fits naturally into a double complex \cite{DoubleComplex}}
\beq{\bm \omega}(\vec  X) =d {\bm \phi}(\vec  X)\label{eq:omega=dphi} \eeq
for some $(n-1)$-form ${\bm \phi}(\vec  X)$. If $\Sigma$ is compact without boundary,
\[{\bm \Omega}[\vec X]=\int_\Sigma{\bm \omega}(\vec  X)=0\]
when one of the variations is a diffeomorphism.
Under the projection ${\cal S} \rightarrow \widehat {\cal S}$ the symplectic form $\widehat {\bm \Omega}$ on $\widehat {\cal S}$
then pulls back to the pre-symplectic form ${\bm \Omega}$ on ${\cal S}$,
\cite{C-W,Lee+Wald}.

If $\Sigma$ has a boundary $\partial \Sigma$
then we can use (\ref{eq:omega=dphi}) to deduce that
\[ {\bm \Omega}[\vec X]=\int_\Sigma {\bm \omega}(\vec X)= \int_{\partial \Sigma} {\bm \phi}(\vec X).\]
Provided $\int_{\partial \Sigma} {\bm \phi}(\vec X)$ vanishes whenever one of the field variations is due to a diffeomorphism then ${\bm \Omega}$ is again a genuine pre-symplectic form. This will be the case for example if the vector field
$\vec  X$ generating the diffeomorphism vanishes fast enough on $\partial\Sigma$. 

Furthermore if the diffeomorphism $\vec  X=\vec K$ corresponds to a Killing symmetry of the solution then $\bm {\omega}(\vec K)=d{\bm \phi}(\vec K)$ vanishes identically and 
$\int_{\partial \Sigma} {\bm \phi}(\vec K)=0$, even when $\vec K$ does not vanish on the boundary \cite{Iyer+Wald}.  If $\partial \Sigma$ consists of two disconnected pieces, $\partial \Sigma = \partial \Sigma_1 \cup \partial \Sigma_2$, then the integral over each piece 
must be equal and opposite and they cancel.  With suitable orientations
\[ {\bm \Phi}[\vec K] = \int_{\partial\Sigma_1} {\bm \phi}(\vec K)=\int_{\partial\Sigma_2} {\bm \phi}(\vec K)\]
can be non-zero.  If one of the pieces, for example $\partial\Sigma_1$, 
is held fixed then $\bm \Phi[\vec K]$ evaluated on $\partial \Sigma_2$
is independent of the $(n-1)$-dimensional surface $\partial\Sigma_2$. 
$\Sigma$ does not have to be a Cauchy surface for this statement to be true.  

\begin{figure}[h]\label{PhiAnnulus}
\centerline{\includegraphics[scale=0.7]{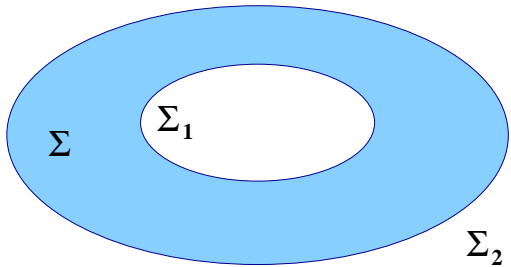}}
\caption{When $\partial \Sigma = \partial \Sigma_2 \cup \partial \Sigma_1$, and
  $\vec K$ is Killing, ${\bm \Phi}[{\vec K}]$ is 
independent of which connected piece of the boundary it is evaluated on,
$\partial\Sigma_1$ or $\partial\Sigma_2$. ${\bm \Phi}[\vec K]$ depends on the fields and a variation thereof but if one segment of the boundary, $\partial\Sigma_1$ say, is fixed we can distort $\partial\Sigma_2$ and move it around, provided it does not pass through a singularity in the geometry ${\bm\Phi}[\vec K]$ evaluated on $\Sigma_2$ does not change.} 
\end{figure}

If
${\bm \Phi}[\vec K]$ is ${\bm \delta}$-exact, and only if it is ${\bm \delta}$-exact, then
\[{\bm \Phi}[\vec K] = {\bm \delta} \bm {{\cal Q}}  [\vec K]\]
for some ${\bm {{\cal Q}}}[\vec K]$.
This ${\bm {{\cal Q}}}[\vec K]$ is a charge related to a Hamiltonian associated with the flow generated by $\vec K$, \cite{Wald2}.

Among the key ingredients to generate conserved quantities from this construction are a general co-ordinate invariant action and a solution of the equations of motion with a Killing vector generating the symmetry.  This formalism was shown in \cite{Iyer+Wald} to reproduce the ADM mass \cite{ADM} for stationary asymptotically flat black holes in Einstein gravity.  At the same time it clarifies the origin of the mysterious factor of two that is well known to arise when comparing the Komar mass with the ADM mass \cite{Wald2,Komar,Katz}.
The construction in \cite{Lee+Wald} is general enough to include theories with a cosmological constant $\Lambda$, of either sign when $\Sigma$ is compact without boundary. When $\Sigma$ has a boundary we can restrict to negative $\Lambda$ so that the asymptotic regime of a black hole solution is well defined and it was shown in \cite{Hajian+Jabbari,Skenderis} that Lee and Wald's formalism correctly reproduces the Henneaux-Teitelboim mass \cite{Henneaux-Teitelboim,GPP} for an asymptotically AdS Kerr black hole.

In this work we shall explore the formalism of Lee and Wald\cite{Lee+Wald} and Wald \cite{Wald2} in some detail and show how it relates to some other definitions of mass in the literature, not just the ADM mass but also the Brown-York mass \cite{Brown+York} and the Bondi 
mass \cite{Bondi} in stationary space-times.  
For example we shall see that the Brown-York mass, which is 
expressed as the difference of two extrinsic curvatures, can be viewed as a 1-form on ${\cal S}$.  The formalism also reproduces the Bondi mass when applied to asymptotically flat space-times using Bondi-Sachs co-ordinates. 

The general formalism has a very nice mathematical structure of a double complex\cite{Bott+Tu}, which is the natural mathematical language for describing cohomology.
This formal structure is described elsewhere \cite{DoubleComplex}.

Section \S\ref{sec:EG} reviews Wald's general construction in the context of Einstein gravity with a cosmological constant. The exposition is given in terms of differential forms on ${\cal M}$ and ${\cal S}$ as this is the most natural framework for treating differential cohomology. 
In \S\ref{sec:Darboux} the symplectic 2-form is derived for non-zero $\Lambda$ and shown to be of Darboux form, extending the results of \cite{Lee+Wald} to include a cosmological constant.
Section \S\ref{sec:TKM}, in which conserved quantities associated with a time-like Killing vector are discussed, contains our main results.
We derive an exact result for the variation of the mass in Einstein gravity,
valid at any distance outside a stationary gravitating mass. 
This includes the asymptotically flat case in \S\ref{sec:BH} 
where the expression simplifies and asymptotically reproduces the ADM mass, as derived in \cite{Iyer+Wald}.  We also relate Wald's expression to the extrinsic curvature of $\partial\Sigma$ and the Brown-York mass \cite{Brown+York}.
It usually stated that the ADM mass in asymptotically flat space-times is completely equivalent to that defined by Brown and York, and a proof is given in \cite{Hawking+Horowitz}, in the formalism presented here these two masses are only the same if the fall-off conditions on the metric are slightly stronger than those usually assumed.  The Bondi mass is also derived using Wald's formalism.

Finally the conclusions are summarised in \S\ref{sec:Conclusions}.
Some more technical details are relegated to a number of appendices.

\section{Einstein gravity \label{sec:EG}}

We shall focus on the Einstein action ${\cal A}$ with a cosmological constant on a space-time ${\cal M}$. We keep the dimension of space, $n$, general
for the moment and will specialise to $n=3$ later.
In units with $G=c=1$ the Lagrangian density is
\beq L = \frac{1}{16 \pi} (R_{a b} \wedge *e^{a b} - 2 \Lambda *1),\label{eq:L}\eeq
where 
\beq R_{a b} = d\omega_{a b} + \omega_{a c}\wedge \omega^c{}_b \label{Rdomega}\eeq
are the curvature 2-forms, $e^a$ are orthonormal 1-forms (vielbeins), $e^{a b}=e^a\wedge e^b$ denotes the wedge product and $*$ is the Hodge star operator. In second order formulation the connection 1-forms $\omega_{a b}$ are determined in terms of the orthonormal 1-forms using the torsion free condition
\[ D e^a= d e^a + \omega_{a b}\wedge e^b=0.\] 
The corresponding action is
\beq  {\cal A}=\frac{1}{16\pi} \int_{\cal M}\bigl(R_{ab}\wedge *e^{ab}-2 \Lambda *1\bigr), \label{AEinstein}\eeq
and the equations of motion are 
\[  {E }^c=\frac{1}{16\pi}\bigl(R_{ab} \wedge * e^{a b c} -2 \Lambda *e^c\bigr)=0,\]
equivalent to
\beq {\cal R}_{a b} = \Lambda \eta_{a b},\label{EoM_E} \eeq
where ${\cal R}_{a b}$  are the components of the Ricci tensor in an orthonormal basis and $\eta_{a b}=\hbox{diag}(-1,+1,\cdots,+1)$.

Under an infinitesimal variation  $e^a \rightarrow e^a + {\bm \delta} e^a$
the linearised equations of motion, with constant $\Lambda$, are 
\[{\bm \delta} {\cal R}_{a b}=0\]
and the variation in the Lagrangian density is
\beq {\bm \delta} L= d {\bm \theta}\label{eq:deltaL}\eeq
with
\beq  d{\bm \theta} = \frac{1}{16\pi}({\bm \delta} R_{ab})\wedge * e^{ab}
=\frac{1}{16\pi}({\bm \delta}{\cal R}) * 1,\label{LEoM} \eeq
where the equations of motion have been used and ${\cal R}={\cal R}^a{}_a$ is the Ricci scalar.
We shall refer to field configurations that satisfy the equations of motion 
together with variations that satisfy the linearised equations of motion as
\lq\lq on-shell''.

Now
\[  ({\bm \delta} R_{ab})\wedge * e^{ab}= D ({\bm \delta}\omega_{ab})\wedge * e^{ab}=
d({\bm \delta}\omega_{ab}\wedge *e^{ab})\]
so
\beq  {\bm \theta} = \frac{1}{16\pi}({\bm \delta}\omega_{ab}) \wedge *e^{ab} 
\mod d\label{Thetad}\eeq
and
\beq
{\bm \omega} = -\frac{1}{16\pi} ({\bm \delta} \omega_{a b}) {\bm \barwedge} 
({\bm \delta} * e^{a b}) \mod d 
\eeq
(the symbol ${\bm \barwedge}$ here represents both the wedge product on ${\cal M}$ and on ${\cal S}$
simultaneously, we hope that the distinction between the pre-symplectic density 
$\bm \omega$ and the connection 1-forms $\omega_{a b}$ is clear).

The variation ${\bm \delta} e^a$ can itself be expanded in the orthonormal basis as
\beq  {\bm \delta} e^a = {\bm \delta}(e^a{}_\mu d x^\mu)=({\bm \delta} e^a{}_\mu) d x^\mu :={\bm \Delta}^a{}_b\, e^b,\eeq
where $x^\mu$, $\mu=0,1,\ldots,n$, are co-ordinates on ${\cal M}$ and 
\beq
{\bm \Delta}^a{}_b= ({\bm \delta} e^a{}_\mu )\bigl( e^{-1}\bigr)^\mu{}_b\label{delta-slash}
\eeq
is a $(n+1)\times (n+1)$ matrix.
Not all such variations actually correspond to changing the metric. Decomposing 
\beq {\bm \Delta}_{ab} = \eta_{ac} {\bm \Delta}^c{}_b\eeq
into symmetric and anti-symmetric parts
\beq  {\bm \Delta}_{ab} = {\bm S}_{ab}  + {\bm A}_{ab}, \label{omegaSA}\eeq with
${\bm S}_{ab}={\bm S}_{ba}$ and
${\bm A}_{ab}=-{\bm A}_{ba}$,
only ${\bm S}_{ab}$ can change the metric, ${\bm A}_{ab}$ merely generate local Lorentz transformations\footnote{From now on we shall use the term \lq gauge transformations' for such local Lorentz transformations, as distinct from diffeomorphisms.} under which $L$ is invariant.

Furthermore not all $\wS_{a b}$ correspond to real changes in the metric, under a diffeomorphism $\vec  X$ 
\beq {\bm S}_{a b} = \frac{1}{2}\big(D_a {\bm  X}_b + D_b {\bm  X}_a\big).\label{diffeo-S}\eeq

As ${\bm \delta}\omega_{ab}$ is linear in ${\bm \delta} e^a$ the decomposition (\ref{omegaSA}) implies a similar decomposition for ${\bm \delta} \omega_{ab}$.
Using the torsion free condition (\ref{app:zeroT})
\[ {\bm \delta}\omega_{ab}= (D_b \wS_{a c} - D_a \wS_{b c} -D_c {\bm A}_{ a b} )e^c.\]
This means that
 $\big({\bm \delta} \omega_{ab}\big)\wedge *e^{ab}$ is not gauge invariant,
however
\[ (D{\bm A}_{ab})\wedge*e^{ab}=d({\bm A}_{ab}*e^{ab})
=d*(e^a \wedge {\bm \delta} e_a)\]
is $d$-exact and we can use the arbitrariness in (\ref{Thetad}) to define
\beq  {\bm \theta}(e^a,{\bm \delta} e^a) 
= \frac{1}{16\pi}\left\{({\bm \delta}\omega_{ab})\wedge*e^{ab} +
d*(e^a \wedge {\bm \delta} e_a)\right\}\label{thetadef1}\eeq
which is gauge invariant by construction.
In terms of $\wS_{ab}$ and the co-variant derivative $D_a$
\beq {\bm \theta}(e^a,{\bm \delta} e^a)=\frac 1 {8\pi}\bigl( D_b \wS_a{}^b - \partial_a \wS_b{}^b\bigr)*e^a.\label{Thetadeltae1}
\eeq

The explicit form of ${\bm \omega}(e^e,\delta_1 e^a,\delta_2 e^a)$ in terms of $(\Delta_1)^a{}_b=\delta_1 e^a{}_\mu (e^{-1})^\mu{}_b$ and
$(\Delta_2)^a{}_b=\delta_2 e^a{}_\mu (e^{-1})^\mu{}_b$ is not very illuminating but for completeness is given in appendix \ref{app:omega}.
Here we just remark that, since it is gauge invariant, 
it only depends on the symmetric variations $({\bm S}_1)_{a b}=\frac{1}{2}\big\{ ({\bm \Delta}_1)_{a b} + ({\bm \Delta}_1)_{b a}\big\}$ and    
$({\bm S}_2)_{a b}=\frac{1}{2}\big\{ ({\bm \Delta}_2)_{a b} + ({\bm \Delta}_2)_{b a}\big\}$. 

Note that (\ref{Thetadeltae1}) has no explicit dependence on the cosmological constant $\Lambda$, though there is an implicit dependence when $e^a$ are on-shell. This is to be expected from (\ref{eq:L}) as the cosmological term in the action only involves $e^a$, not their derivatives, and so cannot influence $d{\bm \theta}$ in (\ref{eq:deltaL}).
In particular the presence of a non-zero $\Lambda$ does not affect the symplectic form and the statement by
Lee and Wald in \cite{Lee+Wald} that the symplectic form takes the Darboux form is unchanged when $\Lambda$ is introduced, as we shall see explicitly in \S\ref{sec:Darboux}.

Diffeomorphisms are generated by an infinitesimal
 vector field $\vec{ X}$ 
\beq {\bm \delta} e^a = \bm{{\cal L}}_{\vec{\bm  X}}\, e^a = d \bm{i}_{\vec{\bm  X}}\, e^a +  {\bm i}_{\vec{\bm  X}}\, d e^a\:.\eeq
When the variation is a diffeomorphism one finds, for the Einstein action (\ref{AEinstein}), 
\beq {\bm \theta}(e^a,\bm{{\cal L}}_\vX\, e^a)=-\frac{1}{16\pi}\left(2e^a\wedge*DD{\bm  X}_a +d*d{\bm  X}\right),\eeq
where ${\bm  X}={\bm  X}_a e^a$ is both a 1-form on ${\cal M}$ and on ${\cal S}$.\footnote{In practice we need not take $\vec  X$ to be infinitesimal.
Since all subsequent formulae are linear in $\vec  X$ we can re-scale $\vec  X \rightarrow \epsilon \,\vec  X$, with $\epsilon \ll 1$ and $\epsilon$ is just an overall factor in all formulae. Indeed we can let ${\bm \epsilon}$ represent a constant 1-form on ${\cal S}$, so that ${\bm \delta} {\bm \epsilon} = - {\bm \epsilon} {\bm \delta}$
and $\vec {\bm X}  = {\bm \epsilon} \vec X$ is a vector on ${\cal M}$ and a 1-form on ${\cal S}$, with ${\bm i}_{\vec \vX}={\bm \epsilon} i_{\vec X}$, \cite{DoubleComplex}.}
Now $DD{\bm  X}_a = R_{ab} {\bm  X}^b$ leading to
\[ {\bm \theta}(e^a,\bm {{\cal L}}_\vX\, e^a) = \frac{1}{16\pi}\left(2\,{\cal R}_{ab}\,{\bm  X}^a *e^b - d*d{\bm  X}\right)= 
\frac{1}{16\pi}\left(2 \Lambda * {\bm  X} - d*d{\bm  X}\right)
\]
where the equations of motion have been used in the last step.
But on-shell
\[L= \frac{\Lambda }{8\pi} * 1\] 
 and by definition $*X = i_{\vec X} *1$, so 
\[ {\bm \theta}(e^a,\bm {{\cal L}}_\vX\, e^a) = {\bm i}_{\vX}\,L + {\bm \WJ}\big( \vec  {\bm X} \big)\]
where
\[ {\bm \WJ}\big( \vec  {\bm X} \big)= - \frac{1}{16\pi} d*d{\bm  X}\]
and  ${\bm \WJ}\big( \vec  {\bm X} \big) = d{\bm \WQ}$ is $d$-exact,  with
\beq {\bm \WQ} = -
\frac{1}{16\pi} * d {\bm  X}.\label{eq:q-def}\eeq
$d\bm\WQ$ is a 2-form on ${\cal M}$. The similarity between
$d\bm\WQ = d*d {\vec X}$ with the vacuum Maxwell equations was observed in
\cite{Simon}.

The symplectic density is obtained from
\bean {
\bm \omega}(\vec  {\bm X}\,) &=& \bdelta {\bm \theta}(e^a,\bm {{\cal L}}_\vX\, e^a)
+ \bm{{\cal L}}_\vX\, {\bm \theta}(e^a,\bdelta e^a)\\
&=& \bdelta \left({\bm i}_{\vX} L - \frac{1}{16\pi} d* d{\bm  X}\,\right)  
+ (d{\bm i}_\vX \,+ {\bm i}_\vX\,d)  {\bm \theta}(e^a,\bdelta e^a)  \\
&=& = d\left(  - \frac{1}{16\pi} \bdelta(* d{\bm  X}\,) +  {\bm i}_\vX\,  {\bm \theta}(e^a,\bdelta e^a) \right),\eean 
where we have used
\[ \bdelta {\bm i}_{\vX} L = - {\bm i}_{\vX} \,\bdelta L
= -  {\bm i}_{\vX} \,d  {\bm \theta}(e^a,\bdelta e^a).\]  
We have proven that
\[{\bm \omega}(e^a,{{\cal L}}_{\vX}\,e^a,{\bm \delta} e^a)= d {\bm\phi}(\vec  {\bm X}\,) \]
with
\beq {\bm \phi}({\bm\vX})=
-\frac{1}{16\pi}{\bm \delta} * d {\bm  X}  +  {\bm i}_{\vX}\, {\bm \theta}(e^a,{\bm \delta} e^a)\eeq    
and
\beq
{\bm \Omega}[\vec {\bm X}] = \int_\Sigma {\bm \omega} (\vX)
=\int_{\partial \Sigma}{\bm \phi}(\vX\,)\label{vxiOmega}\eeq
where ${\bm \Omega}[\vec {\bm X}]={\bm \Omega}[e^a,\bm {{\cal L}}_{\vec {\bm X}}\,e^a,{\bm \delta} e^a]$.
When $\Sigma$ has no boundary ${\bm \Omega}[\vec {\bm X}]$ vanishes and a general ${\bm \Omega}$ will be a genuine pull back from $\widehat {\bm \Omega}$ on $\widehat{\cal S}$.
 If $\Sigma$ has a boundary this is still the case provided ${\vec  X}$ is constrained to vanish on the boundary.\footnote{If $\partial\Sigma$ is some asymptotic regime with unbounded area then this statement must be qualified,
 ${\vec  X}$ must vanish \lq\lq fast enough'' on the boundary.}

If $\vec X=\vec K$ is Killing then $\wS_{a b}=0$, and not only does ${\bm \Omega}[\vec {\bm K}]$ vanish but ${\bm \omega}(\vec {\bm K})$ is identically zero, see (\ref{omegaedede}), independently of any boundary conditions or choice of hypersurfaces. 
This is an important observation: on-shell ${\bm \omega}(e^a,\bm {{\cal L}}_{\vec {\bm K}}\, e^a, {\bm \delta} e^a)={\bm \omega}(\vec {\bm K})$ vanishes identically for any Killing vector $\vec K$. 
Since ${\bm \omega}(\vX)=d{\bm \phi}(\vX)$ this implies that
\beq{\bm \phi}(e^a,\vec{\bm K},{\bm \delta}e^a)=-\frac{1}{16 \pi}{\bm \delta}  * d {\bm K}  +{\bm i}_{\vec {\bm K}}\, {\bm \theta}(e^a,{\bm \delta} e^a)\label{PhiK}\eeq
is $d$-closed, but it need not vanish and can carry useful information.  

If $\vec K$ is purely tangential to $\partial \Sigma$, for example if $\vec K$ generates rotations about the origin and $\partial \Sigma$ is the sphere at infinity, then 
\[ \int_{\partial \Sigma} {\bm i}_{\vec {\bm K}} {\bm \theta} =0\]
and 
\[ 0={\bm \Omega}[\vec {\bm K}] =   -{\bm \delta}\left(\frac{1}{16\pi}  \int_{\partial \Sigma}* d {\bm K}\right) \]
so
\[\frac{1}{16\pi}\int_{\partial \Sigma}  * d {\bm K}\]
is invariant under on-shell perturbations of the metric.
If the boundary $\partial \Sigma$ consists of
disconnected pieces, for example if $\partial \Sigma=\partial \Sigma_1 \cup \partial \Sigma_2$ consists of two separate pieces $\partial \Sigma_1$ and $\partial \Sigma_2$, then define
 \beq {\bm \Phi}_p[{\vec {\bm K}}]   = -\frac{1}{16\pi}\int_{\partial \Sigma_p} {\bm \delta} *  d {\bm K}
\label{Phidef_Komar}
\eeq
($p=1,2$) and, with appropriate orientations, 
\[{\bm \Omega}[\vec {\bm K}]
= {\bm \Phi}_2[{\vec {\bm K}},{\bm \delta} e^a] - {\bm \Phi}_1[{\vec {\bm K}},{\bm \delta} e^a]=0.\]
We can deform $\Sigma$ and move either of the boundaries around,
keeping the other fixed --
provided we do not pass through any singularities in the geometry and
\beq \label{Phi12} {\bm \Phi}[{\vec {\bm K}},{\bm \delta} e^a]
:={\bm \Phi}_1[{\vec {\bm K}}a,{\bm \delta} e^a]
 ={\bm \Phi}_2[{\vec {\bm K}},{\bm \delta} e^a]\eeq
is unchanged by such deformations. 
This statement is independent of any symplectic structure and $\Sigma$ need not be a Cauchy surface. ${\bm \Phi}$ in (\ref{Phi12}) will not change as the boundary of $\Sigma$ is moved around, as long as it does not move through a region containing matter or a singularity in the geometry. 
\[\bm Q[{\vec {\bm K}}]=-\frac{1}{16\pi}\int_{\partial \Sigma_1}  * d \bm K
=-\frac{1}{16\pi}\int_{\partial \Sigma_2}  * d \bm K\]
is the integral of the Komar 2-form associated with the Killing vector ${\vec K}$ \cite{Komar}, it is not itself invariant under metric perturbations of course.

When ${\vec K}$ is not purely tangential to $\partial \Sigma$ however
${\bm \theta}$ can contribute to ${\bm \phi}$ and the story is more involved, but we can still define a charge if
${\bm i}_{\vec {\bm K}} {\bm \theta}$ is $\bdelta$-exact.  
In the most general case we have 
 \beq {\bm \Phi}_p[{\vec {\bm K}}]   = -\frac{1}{16\pi}\int_{\partial \Sigma_p} {\bm \delta} *  d {\bm K}
   +
\int_{\partial \Sigma_p}{\bm i}_{\vec {\bm K}}\, {\bm \theta}(e^a,{\bm \delta} e^a)\label{Phidef_a}
\eeq
If furthermore ${\bm i}_{\vec {\bm K}} {\bm \theta}$ is $\bdelta$-exact, so
${\bm \phi}={\bm \delta}{\bm \Nq}(\vec{\bm K})$ for some $\bm \Nq$, then
\beq{\bm \NQ }[\vec{\bm K}] =\int_{\partial \Sigma_1} {\bm \Nq}(\vec{\bm K})
 =\int_{\partial \Sigma_2} {\bm \Nq}(\vec{\bm K})\label{eq:Q-def}\eeq 
is a candidate for a Noether charge\footnote{This is not in general the same as the Noether charge associated the entropy, as defined in \cite{Wald2}.} associated with the symmetry generated by ${\vec K}$.
${\bm \NQ }[\vec{\bm K}]$ and ${\bm \Nq}(\vec{\bm K})$ 
are 1-forms on ${\cal S}$ but only through their dependence on ${\bm K}$,
they do not themselves involve a metric variation but also are not invariant
under genuine metric perturbations.
Examples are given in \S\ref{sec:BH} for ${\vec K}=\frac{\partial}{\partial t}$ a time-like Killing vector, in which case $\NQ\left[\frac{\partial}{\partial t}\right]$ is a mass.
 
\section{The symplectic form \label{sec:Darboux}}  

We shall now explicitly calculate the symplectic form for Einstein gravity with
a cosmological constant $\Lambda$. The case $\Lambda=0$ was analysed in
\cite{C-W,Lee+Wald,Iyer+Wald}.

Assume the space-time ${\cal M}$ can be foliated into space-like hypersurfaces $\Sigma_t$ of constant $t$. 
Defining lapse and shift functions $N$ and $N^\alpha$ for the foliation in the usual way
we can choose the orthonormal 1-forms\footnote{Greek indices near the middle of the alphabet, $\mu,\nu,\ldots = 0,1,\ldots ,n$, label co-ordinates on ${\cal M}$ while indices $\alpha,\beta,\ldots = 1,\ldots,n$, near the beginning of the alphabet label co-ordinates on $\Sigma_t$. Roman letters $a,b,..=0,1,\ldots,n$ are orthonormal indices on ${\cal M}$ and $i,j,\ldots=1,\ldots,n$ are orthonormal indices for 1-forms $\tilde e^i$ on $\Sigma_t$.}
\[e^a = e^a{}_\mu d x^\mu\]
to decompose as  
\beq  e^0=N dt \qquad\hbox{and}\qquad e^i=\tilde e^i + \frac{N^i}{N}  e^0,
\label{eetilde}\eeq
where
\[\tilde e^i=\tilde e^i{}_\alpha dx^\alpha\]
and $N^i=\tilde e^i{}_\alpha N^\alpha$.
With (\ref{eetilde}) we have made a partial choice of gauge
\beq e^a{}_\mu = 
\begin{pmatrix}
  N & 0 \\ N^i & \tilde e^i{}_\beta
  \end{pmatrix},
\label{Vierbeins}
\eeq
often referred to as the time gauge, and this will be used in the following.

The time-like unit 1-form $n=-e^0$ vanishes on $\Sigma_t$ and the future-pointing unit vector normal to $\Sigma_t$ is
\beq \vec n = \frac{1}{N} ( \partial_t - N^\alpha \partial_\alpha).\label{e0dt}\eeq
In the time gauge (\ref{Vierbeins}) the extrinsic curvature of $\Sigma_t$ 
in ${\cal M}$ takes the form
\beq
\kappa_{a b} = 
\begin{pmatrix}
0 & 0 \\ 0 & \kappa_{i j}
\end{pmatrix}.
\eeq
$\kappa_{i j}$  can be expressed in terms of the time evolution of the dreibein
\[\tau_{i j} := \big(\partial_t \tilde e_{i \alpha}\big)\big(\tilde e^{-1}\big)^\alpha{}_j\]
and the shear of the shift function
\[ \sigma_{\{i j\}} := \frac{1}{2}(\wD_i N_j  + \wD_j N_i ),\]
with $\wD_i$ the 3-dimensional co-variant derivative on $\Sigma_t$,
\[\wD_i N_j = \partial_i N_j +\widetilde \omega_{j k,i} N^k.  \]
In terms of these the extrinsic curvature is
\beq \kappa_{i j} = \frac{1}{N}\big(\tau_{\{i j\}} - \sigma_{\{i j\}}\big),\label{kappasdot}\eeq
with $\tau_{\{ i j \}}=\frac{1}{2}(\tau_{i j} + \tau_{j i})$.

In the time gauge the Einstein Lagrangian takes the well-known form 
\[ L = \frac{1}{16\pi}\big(R_{a b} \wedge *e^{a b} - 2\Lambda * 1\big)
= \frac{1}{16\pi}\big (\kappa_{i j} \kappa^{i j} - \kappa^2 + \widetilde {\cal R} - 2\Lambda)*1 \mod d,\]
with $\kappa$ the trace of $\kappa_{i j}$ and $\widetilde {\cal R}$
is the 3-dimensional Ricci scalar associated with $\tilde e^i=\tilde e^i{}_\alpha d x^\alpha$.
Discarding surface terms $\dot {\tilde e}^i{}_\alpha$ only appears here in $\kappa_{i j}$, through the
$\tau_{\{i j\}}$ term in (\ref{kappasdot}), and
\[\frac{\delta \kappa^{i j}}{\delta \dot{\tilde e}^k{}_\alpha} =
\frac{1}{N} \frac{\delta \tau^{\{i j\}}}{\delta \dot{\tilde e}^k{}_a} 
=\frac{1}{ 2N} \big\{ (\tilde e^{-1})^{\alpha j} \delta^i_k +  (\tilde e^{-1})^{\alpha i} \delta^j_k \big\}.\]
The momentum canonically conjugate to $\tilde e^i{}_\alpha$ in the Hamilton
formulation is the $(n+1)$-form
\[\widetilde \Pi^\alpha{}_i = \frac{\delta L}{\delta \dot {\tilde e}^i{}_\alpha}\]
in terms of which
\[ \widetilde \Pi_{i j} := \tilde e^i{}_\alpha \widetilde \Pi^\alpha{}_i
= \frac{1}{8\pi N}(\kappa_{i j} - \delta_{i j} \kappa)*1 = 
 \frac{1}{8\pi}(\kappa_{i j} - \delta_{i j} \kappa)(dt\wedge \tilde *1),\] 
where $\tilde *$ is the Hodge-$*$ on $\Sigma_t$, with $\tilde * 1 = \tilde e^{1\cdots n}$.
We therefore define momentum $n$-forms on $\Sigma_t$
\bea \widetilde \Pi^\alpha{}_j = (\tilde e^{-1})^{\alpha j} \widetilde \Pi_{ij} 
\qquad\hbox{with}\qquad  \widetilde \Pi_{i j} =   
 \frac{1}{8\pi}(\kappa_{i j} - \delta_{i j} \kappa) \tilde *1.\label{Pitildedef}\eea
 
In the time gauge variations of the metric induce
\beq
{\bm \Delta}^a{}_b =
 {\renewcommand*{\arraystretch}{1.5}\begin{pmatrix}
\frac{{\bm \delta} N}{N} & 0 \\
{\bm \Delta}^i & {\bm \Delta}^i{}_j
\end{pmatrix}}
\eeq
with 
\[{\bm \Delta}^i = {\bm \Delta}^i{}_0 = \frac{1}{N}\big( {\bm \delta} N^i - {\bm \Delta}^i{}_j N^j \big) 
\qquad \hbox{and} \qquad {\bm \Delta}^i{}_j = \big({\bm \delta }\tilde e^i{}_\alpha\big) \big(\tilde e^{-1}\big)^\alpha{}_j.\]

The symplectic structure associated with the action (\ref{AEinstein}) was evaluated in \cite{Lee+Wald} for $\Lambda=0$
and the result is the same for non-zero $\Lambda$.  
${\bm \theta}$ for the action (\ref{AEinstein}) is
\[ {\bm \theta} = \frac{1}{16\pi} \big\{ {\bm \delta}(\omega_{a b} \wedge *e^{a b})  - \omega_{a b} \wedge {\bm \delta} (* e^{a b}) + d *(e^a\wedge {\bm \delta} e_a)\big\}\]
and, using the expressions for $\omega_{a b}$ in the time-gauge given in appendix \S\ref{app:foliation} equation (\ref{omegaij}), this gives
\beq
{\bm \Theta}
 =-\int_{\Sigma_t} \widetilde \Pi^\alpha{}_i {\bm \delta} \tilde e^i{}_\alpha
  -\frac{1}{8\pi} {\bm \delta} \left(\int_{\Sigma_t}  \kappa * e^0\right) 
  - \frac{1}{16\pi}\int_{\partial \Sigma_t}  {\bm \Delta}_i *e^{0 i},
\label{ThetaPi}\eeq
where $\widetilde \Pi^\alpha{}_i {\bm \delta} \tilde e^i{}_\alpha=
\widetilde \Pi_{i j} \wS^{i j}$.
In terms of the extrinsic curvature of $\Sigma_t$ this is
\beq
{\bm \Theta}
 =-\frac{1}{8\pi}\int_{\Sigma_t} (\kappa_{i j} \wS^{i j} + {\bm \delta} \kappa) * e^0
  - \frac{1}{16\pi}\int_{\partial \Sigma_t}  {\bm \Delta}_i *e^{0 i}.
\eeq

From (\ref{ThetaPi}) the pre-symplectic form is
\bea
{\bm \Omega} &=&
\int_{\Sigma_t}  {\bm \delta} \tilde e^i{}_\alpha {\bm \barwedge}\, {\bm \delta}\widetilde \Pi^\alpha{}_i 
 - \frac{1}{16\pi}{\bm \delta} \left(\int_{\partial\Sigma_t} {\bm \Delta}_i * e^{ 0i}\right)
\label{OmegaDarboux}.\nonumber\eea
When the surface term vanishes this is of the Darboux form \cite{Lee+Wald}, the inclusion of a cosmological constant does not change this conclusion.

\section{Time-like Killing vectors and mass \label{sec:TKM}}
 
The Noether charge associated the symmetry of a time-like Killing vector
is of course a mass.  Suppose an asymptotically flat space-time ${\cal M}$ 
is endowed with a time-like Killing vector $\vec K = \frac{\partial}{\partial t}$, with the normalisation fixed by demanding that $\vec K$ has unit length 
asymptotically.  In the time-gauge (\ref{e0dt}) $\vec K$ 
has components 
\beq K^a = (N,N^i)\label{deeteeN}.\eeq
It was shown in  \cite{Iyer+Wald}
that ${\bm \Phi}\left[\frac{\bm \partial} {\bm \partial t} \right]$
in (\ref{Phidef_a}) is the variation of the ADM mass.
With $\partial\Sigma = \partial\Sigma_1 \cup  \partial\Sigma_2$
\beq {\bm \Phi}[\vec {\bm K}]   = -\frac{1}{16\pi}\int_{\partial \Sigma_p} {\bm \delta} *  d {\bm K} + \int_{\partial \Sigma_p}i_{\vec {\bm K}}\, {\bm \theta}(e^a,{\bm \delta} e^a)\label{Phidef-forms}\eeq
is independent of $p=1$ or $p=2$.

We shall now drop the boldface notation for forms on 
${\cal S}$ from here on ---- while it can be useful in understanding the general structure it becomes rather ugly when examining the details of specific examples --- and write
\beq \label{Phidef}
\Phi[\vec K]   = 
\frac{1}{16\pi}\int_{\partial \Sigma_p} \delta *  d K
+\int_{\partial \Sigma_p}i_{\vec K}\, \theta (e^a,\delta e^a)\eeq
The change in sign in the first term here is because ${\bm \delta} e^a$ and
$\vec {\bm K}$ anti-commute in (\ref{Phidef-forms}) while
${\delta} e^a$ and $K^a$ in (\ref{Phidef}) are ordinary commuting quantities. 

An exact expression for $\Phi[\vec K]$ in the time-gauge, when $\vec K = \frac{\partial}{\partial t}$ in 
(\ref{deeteeN}), is derived in appendix \ref{app:OmegaPhi}, equation (\ref{Omega_partial_Sigma_b}),
\beq
    \Phi\left[\frac{\partial}{\partial t}\right]=    \label{Phi_partial_Sigma}
\frac{1}{8\pi }\int_{\sigma}
\left\{
N^2\left(\wD_j\bigg(\frac{S_i{}^j}{N}\bigg)
-\partial_i \bigg( \frac{S}{N}\bigg)\right)
+ N^j {X}_{i j}
-N_i \big(S^{j k} \kappa_{j k} + {\delta}\kappa\big) \right\} *e^{0 i},\eeq
where $S_{i j} = \frac 1 2 (\Delta_{i j} + \Delta_{j i})$, $S=S^i{}_i$ is the 
trace of $S_{i j}$, ${\delta\kappa}={\delta\kappa}^i{}_i$ is the trace of $\kappa_{i j}$ and\footnote{The combination
\[{ \delta} \kappa_{i j} 
       + \kappa_{i k} { \Delta}^k{}_j - {\Delta}_{i k} \kappa^k{}_j \]
is gauge invariant and depends only on $S_{i j}$.}
\beq { X}_{i j}=
       { \delta} \kappa_{i j} 
       + \kappa_{i k} { \Delta}^k{}_j - {\Delta}_{i k} \kappa^k{}_j + \kappa_{i j} S.
\label{Xdef} \eeq
$\sigma$ in (\ref{Phi_partial_Sigma}) is either of the components of $\partial \Sigma$, $\partial \Sigma_1$
or $\partial \Sigma_2$.

 \subsection{Asymptotically flat black holes \label{sec:BH}}
 
 The simplest example of the formalism in the previous section is as
 always the Schwarzschild line element
\[ d s^2 = -\left(1 - \frac{2 m}{r}\right) d t^2 +
\frac{1}{\left(1 - \frac{2 m}{r}\right)} d r^2 + r^2 (d \vartheta^2 + \sin \vartheta^2 d\varphi^2)\]
for which $N^i=0$.
Hypersurfaces of the Schwarzschild geometry with constant $t$ are time-like
for $r<2m$ so in defining $\Sigma_t$ we restrict to $r>2m$.

\subsubsection{ADM mass}

For the Schwarzschild geometry in the time-gauge $\kappa_{i j}=0$ and $N^i=0$ so
$e^i=\tilde e^i$ and  equation (\ref{Phi_partial_Sigma}) simplifies to 
\beq \Phi\left[\frac{ \partial}{ {\partial t}}\right]
=\frac{1}{8\pi }\int_\sigma \big\{ N\big( \wD_j (S^j{}_i)  - \partial_i S \big)
+(\partial_i N) S - (\partial_j N) S^j{}_i\Big)
*e^{0i} \label{Asymptotic_Phi}
\eeq
with $N=\sqrt{1 - \frac{2 m}{r}}$.

We can choose
\[  e^1 = \frac{d r}{\sqrt{1 - \frac{2 m}{r}}}, \qquad  e^2 = r d \vartheta, \qquad  e^3=r \sin\vartheta d\varphi\]
with unit normal to $\Sigma_t$
\[ n = - e^0 = -\sqrt{\left(1 - \frac{2 m}{r}\right)}\, d t.\]
The connection 1-forms in this gauge are 
\[\omega_{01}=-\frac{m}{N r^2} e^0,\qquad \omega_{12}=-\frac{N}{r} e^2,\qquad \omega_{13}=-\frac{N}{r} e^3,
\qquad \omega_{23}=\frac{\cot\vartheta}{r} e^2.\]
 
If we vary the metric by varying the mass $m \rightarrow m + { \delta} m$ then
\[ { \delta}  e^1 = \frac{{ \delta} m}{r - 2 m}  e^1, \qquad { \delta}  e^2 = { \delta}  e^3=0,\]
so \[ { \Delta}^i{}_j = S^i{}_j=\begin{pmatrix}
  \frac{{ \delta} m}{r - 2 m} & 0 & 0 \\ 0 & 0 & 0 \\ 0 & 0 & 0 \\ \end{pmatrix}\]
and
\[{ \delta} \kappa_{i j} =0.\] 
Now\footnote{In an orthonormal basis $\partial_i:= (\tilde e^{-1})^\alpha{}_i \partial_\alpha$.}
\[ \wD_j S_i{}^ j - \partial_i S_j{}^j =
\frac{2 { \delta} m}{r^2 \left( 1 - \frac{2 m}{r} \right)^{\frac 1 2}} \delta^1_i
\qquad \hbox{and} \qquad \partial_i N = \frac{m}{r^2 \left( 1 - \frac{2 m}{r} \right)^{\frac 1 2}}\delta^1_i,\]
giving (with $*1=e^{0123}$ and $*e^{01}=e^{2 3}$)
\[ \Phi\left[\frac{ \partial}{{\partial t}}\right] 
= \frac{1}{8 \pi} \int_\sigma \left(\frac{2 { \delta} m}{r^2}\right) r^2 \sin\vartheta\, d\vartheta\, d\varphi
=\frac{{ \delta} m}{4 \pi} \int_\sigma \sin\vartheta\, d\vartheta\, d\varphi.\] 
For example we could take $\Sigma$ to be a thick solid shell with $r_1\le r \le r_2$, then the boundary $\partial\Sigma$ consists of two spheres with radii $r_1$ and $r_2$ (in particular it is not necessary to take $r_2\rightarrow \infty$).  Taking $\sigma$ to be the outer sphere gives
\[\delta {\NQ}=
\Phi =\frac{{ \delta m}}{4\pi} \int_{S^2}\sin\vartheta d\vartheta d\varphi = { \delta} m\]
so $\Phi$ is the variation of the mass parameter, which is therefore identified with the physical mass.
Since $\Phi\big[\frac{ \partial}{{\partial t}}\big]$ 
is independent of $r$ we can
calculate it using whatever value of $r$ is convenient. Indeed we can even smoothly distort $\sigma$
to any arbitrary shape, as long as it encloses the origin and subtends a solid angle of $4\pi$ we will always
get the same answer.\footnote{Of course the value that we get for the mass depends on the normalisation of the Killing vector
and, for asymptotically flat space-times, this is naturally fixed by demanding that $\vec K$ has unit length
when $r\rightarrow \infty$.}
More generally, for any space-time with a stationary metric which is asymptotically flat, 
we can evaluate $\Phi\left[\frac {\partial}{\partial t}\right]$
on a sphere of large $r$ in polar co-ordinates.
Since $\frac{\partial}{\partial t}$ is Killing $\tau_{\{ i j \}}$ vanishes in 
(\ref {kappasdot}) and we can assume that
\beq N \sim 1+ O\left(\frac{1}{r} \right), \quad N_i  \sim O\left(\frac{1}{r} \right), \quad  
\partial_i N  \sim \wD_i N_j \sim \kappa_{i j} \sim O\left(\frac{1}{r^2}\right).\label{flat-fall-off-conditions}\eeq
These conditions include the case of asymptotically flat rotating black holes.  A stationary metric has $\frac{\partial}{\partial t}$ as a Killing vector but we do not assume that the variation ${\delta} e^a$ shares this symmetry,
we can only assume the fall-off 
\[S^i{}_j \sim  O\left(\frac{1}{r} \right)\qquad 
\wD_i S^j{}_k \sim  O\left(\frac{1}{r^2} \right)\qquad
\delta \kappa_{ \{ij\} } \sim O\left(\frac 1 r \right), \]
the last since $\delta \tau_{\{ i j \}}$ could be of order $1/r$ in 
(\ref {kappasdot}).
Here however the linearised equations of motion are invoked and 
satisfying these requires that, for an asymptotically flat metric,  
\[\delta \tau_{\{ i j \}}\sim O\left(\frac{1}{r^2} \right).\]

With these asymptotic conditions, $N^j{X}_{i j}\sim O\left(\frac{1}{r^3}\right)$ and
(\ref{Phi_partial_Sigma}) reduces to 
\beq \Phi\left[\frac{ \partial}{ \partial t}\right]= \frac{1}{8\pi }\int_{S^2}
\big( \wD_j (S_i{}^j)  - \partial_i S \big)  \tilde * \tilde e^i + O\left(\frac{1}{r} \right)\label{delta-Wald-mass}
\eeq
which is the variation of the ADM mass \cite{Iyer+Wald,ADM}. 

 This variation has been calculated here asymptotically as
$r\rightarrow\infty$ but it is stressed that, in principle at least, this is not necessary --- we can smoothly distort the sphere at infinity to any other sphere (provided we do not pass through any matter to reach it by the distortion, otherwise (\ref{AEinstein}) is not the correct action to use) 
and the formalism ensures that we would have obtained the same answer for any
stationary asymptotically flat metric.  It is not necessary to go to asymptotia to evaluate variations in the ADM mass for a stationary space-time. 

\subsubsection{The Brown-York mass}

It is well known that in asymptotically flat space-time the ADM mass \cite{ADM} is related
to the Brown-York mass \cite{Brown+York}, indeed it is usually stated that
they are exactly equivalent. The earliest reference to their equivalence appears
to be \cite{Hawking+Horowitz}. In this section we shall investigate how this 
relates to the Wald formalism of \S\ref{sec:BH}, specifically equation (\ref{delta-Wald-mass}).

The Brown-York mass in asymptotically flat space-time is defined using the extrinsic curvature of the asymptotic boundary of $\Sigma$. If $\tilde n$ is the unit normal to
$\partial\Sigma$ then the extrinsic curvature of $\partial\Sigma$ in $\Sigma$
is
\[ \tilde\kappa_{i j} = \frac{1}{2}\widetilde P_i{}^k \widetilde P_j{}^l  (\wD_{k} \tilde n_{l} + \wD_{l} \tilde n_{k}\big)\] 
where
\[\widetilde P_i{}^{j}=\big(\delta_i^j - \tilde n_i \tilde n^j  \big) \]
projects from the (co-)tangent space of $\Sigma$ onto the (co-)tangent space of $\partial\Sigma$.
The trace $\tilde\kappa=\tilde\kappa_i{}^i$
can be obtained from
\beq \tilde d \,\tilde * \,\tilde n = \tilde\kappa \,\tilde *1\label{tracekappa}\eeq
where $\tilde d = \tilde e^i \partial_i$ is the exterior derivative 
and $\tilde *$ the Hodge duality operator 
on $\Sigma_t$
at constant $t$. 
If $\tilde\kappa_0$ is the trace of the extrinsic curvature of $\partial\Sigma$ with the flat metric on $\Sigma$ then the Brown-York mass \cite{Brown+York} is
\beq
M_{B-Y}=\frac{1}{8\pi} \int_{S^2|_{t,\infty}} (\tilde\kappa_0 - \tilde\kappa) \tilde * \,\tilde n
= -\frac{1}{8\pi} \int_{S^2|_{t,\infty}} (\delta\tilde\kappa) \,\tilde * \,\tilde n.
  \label{Brown-York-mass}
\eeq
This is related to (\ref{delta-Wald-mass}) as follows. 
Under a perturbation of the metric (\ref{tracekappa}) can be used to show that, in the gauge (\ref{Vierbeins}), 
\beq 
\int_\sigma \big\{({ \delta} \tilde \kappa)  +  \tilde\kappa_{ij}  S^{ij}\big\} \tilde * \tilde n
=\int_\sigma 
\bigl( \partial_i S - \wD_j S^j{}_i \bigr)\tilde * \tilde e^i.
\label{eq:B-Y-ADM}\eeq
Hence, at large $r$ in asymptotically flat space-time with the fall-off conditions (\ref{flat-fall-off-conditions}), 
\beq  \Phi\left[ \frac{ \partial}{ {\partial t}}\right] = 
-\frac{1}{8\pi}
\int_{S^2|_{t,r}} \big\{ ({ \delta} \tilde\kappa)+
\tilde\kappa_{ij} S^{ij}\big\}\tilde * \,\tilde n
+ O \left( \frac 1 r \right).\label{Phi-BY}
\eeq
Now let $\delta\tilde \kappa = \tilde \kappa - \tilde \kappa_0$ be the deviation of the trace of the extrinsic
curvature of the asymptotic boundary $S^2$ of $\Sigma$ from its flat space value
\[ \tilde \kappa_0 = \frac{2}{r}.\]
Asymptotically the extrinsic curvature has the form
\[ \tilde \kappa_{i j} = \frac{1}{r} \widetilde P_{i j} + O\left( \frac 1 {r^2} \right).\]
This implies that  $\tilde \kappa_0-\tilde \kappa \sim O\bigl(\frac{1}{r^2}\bigr)$
and it is these $1/r^2$ terms that contain information about the mass since
\[ \frac{1}{r^2}\int_{S^2|_{t,r}}\kern -10 pt \tilde *\, \tilde n = 4\pi + O\left( \frac 1 r \right)\]
and
\beq 
\Phi\left[ \frac{ \partial}{ {\partial t}}\right] = 
\frac 1 {8\pi} \int_{S^2|_{t,\infty}} \kern -10pt 
(\tilde\kappa_0 - \tilde\kappa) \tilde * \,\tilde n  +\lim_{r\rightarrow \infty}
\left(\frac 1 {8\pi r} \int_{S^2|_{t,r}} \kern -10pt S_\perp\,\tilde *\,\tilde n\right),
\eeq
where $S_\perp= \widetilde P_{ij}S^{ij}$ is the transverse trace of $S_{ij}$
(it does not matter whether or not we use the flat metric for $\tilde *\,\tilde n$ in (\ref{Phi-BY}) since the difference is $O\left(\frac{1}{r}\right)$).
We see from (\ref{Brown-York-mass}) that $\Phi$ equals the Brown-York mass if
\[ \lim_{r\rightarrow\infty}  \left(\frac {1} {r}\int_{S^2|_{t,r}} \kern -5pt S_\perp\, \tilde * \,\tilde n\right)=0.
\]
It is not sufficient that $S_{i j}$ falls off like $1/r$, in addition
the transverse trace of the metric perturbation, $S_\perp$, must fall off faster than $1/r$ (for a perturbation corresponding to a gravitational wave moving radially outward this is guaranteed since $S_\perp =0$).

This analysis shows that the the difference $\tilde \kappa_0 - \tilde\kappa$ in 
the definition of the Brown-York mass is best viewed as a 1-form on the
space of solutions ${\cal S}$.

\subsubsection{The Bondi mass\label{sec:BondiMass}}

In Bondi-Sachs co-ordinates $(u,r,\vartheta,\varphi)$, \cite{Bondi-Sachs}, the line element is
\[ d^2 s = -U^2 e^{2 W} d u^2 -2 e^{2 W} d u d r + r^2 h_{\alpha \beta} ( d x^\alpha - V^\alpha d u)( d x^\beta - V^\beta d u)\]  
where $u=t-r$ is a light-like co-ordinate, $r$ is a radial co-ordinate and $x^\alpha=(\vartheta,\varphi)$ are co-ordinates on a 2-sphere with metric components $h_{\alpha\beta}$.
In general $U(u,r,\vartheta,\varphi)$, $W(u,r,\vartheta,\varphi)$ and  $V^\alpha(u,r,\vartheta,\varphi)$ are functions of all four co-ordinates, but we require at least that
\[ U \ \mathop{\longrightarrow}_{r\rightarrow \infty} \ 1,\qquad
W \ \mathop{\longrightarrow}_{r\rightarrow \infty} \ 1,\qquad
V^\alpha  \ \mathop{\longrightarrow}_{r\rightarrow \infty} \ 0.
\] 
With foresight in relation to the Bondi mass it is useful to replace $U$,
without any loss of generality,
with the function $M(u,r,\vartheta,\varphi)$ defined via
\beq U=1-\frac {2 M}{r}\eeq
with $M$ finite as $r\rightarrow \infty$.

We can choose orthonormal 1-forms
\beq e^0= \exp(W)\big(U du + U^{-1} dr\big),
\quad e^1 = U^{-1} \exp(W) d r, \quad e^i = r\tilde e^i - V^i du,
\label{BS-gauge1}\eeq
with  $\tilde e^{i}$ ($i=2,3$) orthonormal 1-forms for the 2-sphere metric 
$h_{\alpha\beta}$ and \[V^i=r\tilde e^i_\alpha V^\alpha.\]
We are free to choose a gauge in which 
\beq \tilde e^i = C^i{}_j \hat e^i \label{C-matrix}\eeq
where $\det C=1$ and 
\[ \hat e^2 = d\vartheta, \qquad \hat e^3 =\sin\vartheta d\varphi\]
are orthonormal 1-forms for the round unit 2-sphere. 
Asymptotically we demand that
$C^i{}_j = \delta^i{}_j + O\left( \frac 1 r \right)$ for large $r$ but the condition $\det C=1$ ensures that volume of the 2-sphere is $4\pi$ for all $r$
(see appendix \ref{app:B-S}).
We shall call (\ref{BS-gauge1}) and (\ref{C-matrix}) the Bondi-Sachs gauge.

Now $\vec K = \frac{\partial}{\partial t} =\frac{\partial}{\partial u}$
has metric dual 1-form
\[ K = - U \exp(W) e^0 -V_i \,e^i \]
and using this one finds that, on a sphere defined by $u$ and $r$ constant,
\bea \frac{1}{16\pi}\int_{S^2} \delta (* d K) &=& \label{deltastarKBondi}\\
&&\kern -70pt\frac{1}{8\pi}\delta \left(\int_{S^2} 
\left\{\frac{1}{2}(U^2)' + U^2 \,W'- \dot W - V^i \partial_i W 
-\frac 1 2 e^{-2W}  V^i V'_i \right\} r^2 \hat e^{23} \right).\nonumber\eea
with $\dot W = \partial_u W$ and $W' = \partial_r W$.
The expression for $ \int_{S^2} i_{\vec K} \theta$ is more complicated but
if we assume that $V^i \sim \frac{1}{r}$ and
$W=1+O\left(\frac 1 r  \right)$, in order to ensure asymptotic flatness,
it takes the asymptotic form 
\beq
\int_{S^2} i_{\vec K} \theta=
\frac{1}{8\pi}\int_{S^2} 
\left\{ \frac{\delta M}{r^2}  +\delta \dot W - \delta W' 
+ \Bigl(1-\frac{2 M}{r}\Bigr)\frac{\delta W}{r} \right\} r^2 \hat e^{2 3}+O\left( \frac 1 r\right)\label{iKThetaBondi}
\eeq

Adding (\ref {deltastarKBondi}) and (\ref{iKThetaBondi}) the $\delta \dot W$ terms
cancel and
\bea \frac{1}{16 \pi} \left(\int_{S^2} \delta (* d K)\right)
+ \int_{S^2} i_{\vec K} \theta&=& \\
&& \kern -180pt
 \frac{1}{8\pi}\int_{S^2}  \Big\{ 
2 (\delta M -M \delta W ) +r \big[ \delta W - \delta M' - 2 \delta (M W')\big]
- r^2 \delta W'  
\Big\}  
\hat e^{2 3} + O\left( \frac 1 r \right).\nonumber
\label{Bondi-Wald1}\eea
 
Demanding that the metric is asymptotically flat imposes the conditions
\bea M(u,r,\vartheta,\varphi) &=& m(u,\vartheta,\varphi) + O\left(\frac{1}{r}\right),\nonumber\\
W(u,r,\vartheta,\varphi) &=& \frac{w(u,\vartheta,\varphi)}{r} + O\left(\frac{1}{r^2}\right),\nonumber
\eea
giving
\beq \frac{1}{16 \pi} \left(\int_{S^2} \delta (* d K)\right)
+ \int_{S^2} i_{\vec K} \theta=
 \frac{1}{4\pi}\int_{S^2}  (\delta m  + \delta w)
\hat e^{2 3} + O\left( \frac 1 r \right).
\label{Bondi-Wald2}\eeq

In general the Bondi mass is defined to be 
\beq M(u) = \frac{1}{4\pi} \int_{S^2} m(u,\vartheta,\varphi) \sin \vartheta d \vartheta d \varphi
\label{Bondi_Mass}\eeq
and here we invoke the linearised equations of motion, at order $\frac{1}{r^3}$ the Einstein
equations actually require that $W\sim \frac 1 {r^2}$ so $\delta W\sim \frac 1 {r^2}$ also and $\delta w=0$.

Finally
\beq \frac{1}{16 \pi} \left(
\int_{S^2} \delta (* d K)\right)  + \int_{S^2} i_{\vec K} \theta
= \delta M + O\left( \frac 1 r \right).\label{Bondi-Wald}\eeq
so Wald's expression indeed equals the variation of the Bondi mass.
Again, since $\vec K$ is killing, the general formalism ensures that any 
value of $r$ could have been used in the calculating the left hand side of (\ref{Bondi-Wald}) and the answer would always be the same.

\section{Conclusions \label{sec:Conclusions}}

 The phase space formulation of a dynamical system is ideally suited to the discussion of conserved quantities and symmetries of relativistic systems which are invariant under diffeomorphisms, such as general relativity,
are no exception.  For general relativity the symplectic form ${ \bm \Omega}$ was derived by Witten and Crnkovi\'c in \cite{C-W} and reformulated by Lee and Wald where it was shown in \cite{Lee+Wald} to have the Darboux form in asymptotically flat space-times.  This conclusion is not changed when a cosmological
constant is included.

For stationary solutions of Einstein's equations,
when $\frac{\partial}{\partial t}$ is Killing,
$\Phi$ in (\ref{Phidef}) is a 2-form on the space of solutions which is
independent of the surface (more generally $(n-1)$-dimensional submanifold) $\sigma$ in ${\cal M}$ on which it is calculated.

 If the solution is that of an asymptotically flat stationary space-time
 $ \Phi$ is the variation of both the ADM mass and the Brown-York mass,
 when $\Sigma$ is space-like
and yields the Bondi mass when $\Sigma$ is an appropriately chosen light-like
hypersurface respectively.
The analysis here lends further support to the
suggestion that Lee and Wald's expression
$ \Phi[e^a,{\cal L}_{\frac{\partial}{\partial t}} e^a,{ \delta} e^a]$ 
in equation (\ref{Phidef}) thus serves to unify the different definitions of mass in general relativity that appear in the literature and is a universal expression for the 
variation of the mass associated with a stationary solution of any diffeomorphism invariant theory.

From a mathematical point of view the construction fits very neatly into a double complex
that captures the cohomology of the various forms involved, details of this mathematical structure are given elsewhere \cite{DoubleComplex}.

\newpage

\appendix

\section{Differential form notation \label{app:forms}}

\subsection{Connection and curvature}

For a given metric let $e^a$ denote a set of associated orthonormal 1-forms (a tetrad in four dimensions). 
Our conventions are that orthonormal indices are raised and lowered with 
\[\eta_{a b}= \eta^{a b} =
\begin{pmatrix} -1 & 0 & 0 & 0 \\ 0 & 1 & 0 & 0 \\ 0 & 0 & 1 & 0 \\ 0 & 0 & 0 & 1 \end{pmatrix}.
\]
When 1-forms are wedged together we use the short-hand notation
\[ e^{a_1 a_2 \cdots a_n} = e^{a_1}\wedge  e^{a_1} \wedge \cdots \wedge e^{a_n}.\]
$i_a$ denotes contraction with the orthonormal vector metric dual to $e^a$ so, for example,
\[ i_a e^{b c} = \delta^b_a\, e^c - \delta^c_a\, e^b.\]
The associated torsion free connection 1-forms can be expressed in terms of the $e^a$ as
\beq\omega_{a b}=\frac 1 2 \bigl(e^c i_a i_b d e_c - i_a d e_b + i_b d e_a \bigr) \label{omega-ide}\eeq
where $d$ is the exterior derivative. The co-variant exterior derivative is denoted $D$, in terms of which 
the torsion free condition is 
\beq D e^a = d e^a + \omega^a{}_b \wedge e^b=0.\label{app:zeroT}\eeq
The curvature 2-forms are
\[ R_{ab} = d \omega_{a b} + \omega_a{}^c \wedge \omega_{c b}=\frac 1 2 R_{a b c d}e^{c d}\]
where $R_{a b c d}$ are the components of the Riemann tensor in the chosen orthonormal basis.
The components of the associated Ricci tensor, ${\cal R}_{a b}$, and the Einstein tensor, $G_{a b}$, can be extracted from
\[ R_{b c}\wedge  *e^{a b c} = \bigl( 2{\cal R}^a{}_b-{\cal R}\delta^a{}_b \bigr)* e^b= -2 G^a{}_b *e^b \]
where ${\cal R}={\cal R}^a{}_a$ is the Ricci scalar 
and $*$ is the Hodge duality operator.

If the metric is varied infinitesimally the orthonormal 1-forms must change,
\[ e^a \rightarrow e^a + {\delta} e^a.\]
Demanding that the connection 1-forms also change so as to preserve the torsion free condition
implies that
\[ {\delta}\bigl (D e^a) = D {\delta} e^a + {\delta} \bigl(\omega^a{}_b\bigr)\wedge  e^b =0 \]
allowing ${\delta}\omega_{ab}$ to be determined from $\omega_{ab}$ and ${\delta} e^a$ through
\beq  {\delta}\omega_{ab} = 
\frac 1 2 \left( e^c i_a i_b\, D{\delta} e_c - i_a D{\delta} e_b + i_b\, D{\delta} e_a\right).\label{omegadeltae}\eeq
The variation in the curvature 2-forms is
\beq  {\delta} R_{ab}=d({\delta}\omega_{ab}) +\omega_a{}^c\wedge {\delta}\omega_{cb} +
\omega_b{}^c \wedge {\delta}\omega_{ac} =D({\delta}\omega)_{ab}.\label{deltaRab}\eeq

\section{Explicit expression for ${ \bm\theta}$ and ${ \bm\omega}$ in Einstein gravity \label{app:omega}}

From (\ref{omegadeltae}), keeping only symmetric variations,
\beq {\theta}(e^a,{\delta} e^a)=\frac 1 {8\pi}\bigl( D_b \ws_a{}^b - \partial_a \ws_b{}^b\bigr)*e^a.\label{app:Thetadeltae1}
\eeq
For completeness we give here the explicit form of ${ \omega}$ under two
variations, ${\delta_1}$ and ${\delta_2}$ with
\[(\ws_1)^a{}_b = ({\delta_1} \tilde e^a{}_\mu) (\tilde e^{-1})^\mu{}_b
\qquad \hbox{and} \qquad
(\ws_2)^a{}_b = ({\delta_2} \tilde e^a{}_\mu) (\tilde e^{-1})^\mu{}_b,
\]
\bea
{\omega}(e^a,{\delta_1} e^a,{\delta_2} e^2)&=&
\frac 1 {8\pi}\Bigl\{(\ws_1)^{bc} D_a (\ws_2)_{b c} - 2 (\ws_1)^{bc} D_c (\ws_2)_{ba} \nonumber \\ 
&&  \hskip 50pt + \ws_1 D_b (\ws_2)^b{}_a
 +(\ws_1)_a{}^b \partial_b \ws_2 - \ws_1 \partial_a \ws_2\Bigr\}*e^a \nonumber \\
&&\hskip 100pt -(1 \leftrightarrow 2), \label{omegaedede} \eea
where $\ws_1 := (\ws_1)_c{}^c$ is the trace of $(\ws_1)_b{}^c$ and similarly for $\ws_2$.
The explicit form of ${
  \omega}$ is 
not very useful but of course it vanishes if either $(\ws_1)_{a b}$ or $(\ws_2)_{a b}$ is zero, in particular this is the case if either of the variations is generated by a Killing symmetry.

\subsection{Space-like foliation \label{app:foliation}}

For a space-time ${\cal M}$ with metric $g_{\mu\nu}$ and co-ordinates $x^\mu$ foliate ${\cal M}$ with constant time hypersurfaces.
Let $x^\mu = (t,x^\alpha)$  where $\alpha=1,2,3$ and $t$ is a time co-ordinate.  
We use the standard ADM decomposition: assume that $t=const$ are space-like hypersurfaces, $\Sigma_t$, and denote the induced metric on $\Sigma_t$ by $h_{\alpha\beta}(t)$. The 4-dimensional line element decomposes as
\bea ds^2 &=& g_{\mu\nu} d x^\mu dx^\nu = g_{tt}dt^2 + 2 g_{t\alpha} dt dx^\alpha + g_{\alpha\beta} dx^\alpha d x^\beta\nonumber \\
&=&-N^2 dt^2 + h_{\alpha\beta} (dx^\alpha + N^\alpha dt)(dx^\beta + N^\beta dt),\eea
where $g_{tt}=-N^2 + h_{\alpha\beta} N^\alpha N^\beta$,
$g_{t\alpha}=g_{\alpha\beta} N^\beta $ and $ h_{\alpha\beta} = g_{\alpha\beta}$. 

The orthonormal 1-forms $e^a$ for the metric $g$ can be expressed in a co-ordinate basis as
\[ e^a=e^a{}_\mu d x^\mu\]
while
\[\tilde e^i=\tilde e^i{}_\alpha dx^\alpha,\]
with $i=1,2,3$, are orthonormal 1-forms for $h$.  
Then $\tilde e^i{}_\alpha = e^i{}_\alpha$ and
\beq  e^0=N dt \qquad\hbox{and}\qquad e^i=\tilde e^i + \frac{N^i}{N}  e^0,
\label{app:eetilde}\eeq
with $N^i=e^i{}_\alpha N^\alpha$ the orthonormal components of the shift vector.
The connection 1-forms on $\Sigma_t$ are defined in the usual way
\[\tilde d\tilde e^i + \widetilde \omega^i{}_j \wedge \tilde e^j =0\]
with $\tilde d= \tilde e^i \partial_i$ the exterior derivative on $\Sigma_t$
at constant $t$.

In this gauge
\bea e^a{}_\mu = \begin{pmatrix}
N & 0 \\ N^i & \tilde e^i{}_\beta
\end{pmatrix}, \qquad
(e^{-1})^\mu{}_a = 
{\renewcommand*{\arraystretch}{1.2}
\begin{pmatrix}
\frac 1 N & 0 \\ 
-\frac{N^\alpha}{N} & (\tilde e^{-1})^\alpha{}_j
\end{pmatrix}}\label{app:Vierbeins}
\eea
and the unit vector normal to $\Sigma_t$, $\vec n$, has orthonormal components $n^a=(1,0,0,0)$ so the metric dual 1-form is
$n=n_a e^a = -e^0$.

Metric variations are described by
\beq 
{\Delta}^a{}_b =
{\renewcommand*{\arraystretch}{1.2}
\begin{pmatrix}
  \frac {{\delta} N} N & 0 \\  
 \frac{ \tilde e^i{}_\alpha {\delta} N^\alpha}{N} &  {\Delta}^i{}_j
\end{pmatrix}},\label{deltae}
\eeq
with $ {\Delta}^i{}_j = ({\delta} \tilde e^i{}_\alpha)(\tilde e^{-1})^\alpha{}_j$.  This can be decomposed into symmetric and anti-symmetric parts
\[\ws_{i j} = \Delta_{\{i j\}}=\frac 1 2 ( {\Delta}_{i j} +  {\Delta}_{j i}), \qquad  {A}_{i j} = \Delta_{[i j]} = \frac 1 2 ( {\Delta}_{i j} -  {\Delta}_{j i}).\]
If we define the shift 1-forms as 
\[\widetilde N=h_{\alpha\beta} N^\alpha d x^\beta=N_i \tilde e^i\] 
then\footnote{\label{footnote:gauge} $N_i \tilde e^i$ is invariant under spatial gauge transformations, ${\Delta}_{i j}= {A}_{i j}$, $({\delta} \widetilde N)_i=0$,
  so ${\delta} N^i = - N^k {A}_{k i}$ under such a gauge transformation.}
\beq{\delta} \widetilde N = ({\delta} \widetilde N)_i e^i
\qquad \hbox{with} \qquad  ({\delta} \widetilde N)_i = {\delta} N_i + N_j  {\Delta}^j{}_i,\label{deltaNi}\eeq
and
\beq {\Delta}^i:={\Delta}^i{}_0 = \frac{\tilde e{}^i{}_\alpha {\delta}N^\alpha}{N}
=\frac{1}{N} ({\delta} N^i - {\Delta}^i{}_j N^j )
= \frac{1}{N} \bigl(({\delta} \widetilde N)^i - 2 \ws^i{}_j N^j\bigr)
.\label{deltaei0}\eeq

The vector $\vec K=\frac{\partial}{\partial t}$ has contravariant components
$ K^\mu=(1,0,0,0)$, so $K^a=(N,N^i)$  and 
and the metric dual 1-form is
\beq   K =-N e^0 + N_i e^i. 
\label{Ke}\eeq
Under the diffeomorphism generated by $\frac{\partial}{\partial t}$
the change in the metric components is $\partial_t g_{\mu\nu}$ and we define
\[ \tau^a{}_b = \dot e^a{}_\mu (e^{-1})^\mu{}_b,\]
where $\dot \ = \partial_t$, 
so
\[ \tau^i{}_0 = \frac{(\tilde e^i{}_\alpha)\dot N^\alpha}{N}\]
and
\[ \dot{\tilde e}^i = \dot{\tilde e}^i{}_\alpha d x^\alpha =  {\tau}^i{}_j \tilde e^j\]
where
\[ {\tau}^i{}_j = \dot{\tilde e}^i{}_\alpha \bigl({\tilde e}^{-1}\bigr)^\alpha{}_j.\]

If $\frac{\partial}{\partial t}=\vec K $ is Killing then
\beq {{\cal L}}_{\vec K}g_{\mu\nu}=\frac{\partial g_{\mu\nu}}{\partial t}:=\dot g_{\mu\nu}=0\eeq
in this basis, so $\dot h_{\alpha\beta}$, $\dot N^\alpha$ and $\dot N$ all vanish. However  
$\dot N^i$ is not necessarily zero, 
only
$\dot N^\alpha=0$ and $\dot e^i{}_0=0$, and
\[\dot N^i = \dot e^i{}_\alpha N^\alpha\]
so the orthonormal triad $\tilde e^i$ are not necessarily time independent, even if the metric $h$ is.
While $\tau_{\{i j \}}$ is zero when $\frac{\partial}{\partial t}$ is Killing
$\tau_{[i j ]}$ need not be, rather $\dot N^\alpha=0$ so
\beq\dot N_i - \tau_{[i j]} N^j =0.\label{Ndotzero}\eeq 

In general the connection 1-forms in the gauge (\ref{app:Vierbeins})
are\footnote{By definition
\beq  
\widetilde \omega_{i j} = 
\widetilde \omega_{i j,k} \tilde e{}^k = \widetilde\omega_{ij,k}\left(e^k- \frac{N^k}{N} e^0\right)\eeq
so
\beq \omega_{ij,0}=  -\frac {1}{N} \left( \sigma_{[ij]} +\tau_{[i j ]}
+ N^k \widetilde \omega_{i j,k}\right). \eeq}
\begin{eqnarray}
\omega_{0i}&=-\frac{\partial_i N} {N} e^0 +
\frac {1} {N} \bigl(\sigma_{\{i j\}} -\tau_{\{i j \}}\bigl) e^j \nonumber\\
\omega_{i j}&=\widetilde \omega_{i j} -  \frac {1}{N}\bigl(\sigma_{[i j]} + \tau_{[i j]}\bigr)e^0, 
\label{omegaijA}\end{eqnarray}
where $\sigma_{\{i j\}}$ and $ \sigma_{[i j]}$ are the shear and
vorticity of the shift vector,
\[\sigma_{\{i j\}}:=\frac 1 2 (\widetilde D_i N_j + \widetilde D_j N_i),\qquad 
\sigma_{[i j]}:=\frac 1 2 (\widetilde D_i N_j - \widetilde D_j N_i),\] 
with
\beq \sigma_{i j}=\widetilde D_i N_j = \partial_i N_j + \widetilde \omega_{j k,i} N^k \eeq
the co-variant derivative of the shift functions on $\Sigma_t$.
 
The extrinsic curvature of $\Sigma_t$ is 
\[\kappa_{a b} = \frac 1 2 (\delta_a{}^c - n_a n^c)(\delta_b{}^d-n_b n^d)(D_c n_d + D_c n_d).\]
In the time-gauge, $n_a=(-1,0,0,0)$, and
\[ D_a n_b = \partial_a n_b + n^c \omega_{b c, a} = -\omega_{b 0,a}\]
so
\beq \kappa_{a b} = \begin{pmatrix}
0 & 0 \\ 0 & \kappa_{i j}\\
\end{pmatrix} \qquad\hbox{with} \qquad
\kappa_{i j} = \frac{1}{2}(\omega_{i 0,j} + \omega_{j 0,i})
=\frac{1}{N}(\tau_{\{i j \}} - \sigma_{\{i j\}}).\label{Sigmakappa}\eeq
We can re-write (\ref{omegaijA}) as
\begin{eqnarray}
\omega_{0 i}&=&-\frac{\partial_i N} {N} e^0 -\kappa_{i j} e^j \label{omega0i}\\
\omega_{i j}&=&\widetilde \omega_{i j} -  \frac {1}{N}\bigl( \sigma_{[i j]} + \tau_{[i j]}\bigr)e^0, 
\label{omegaij}
\end{eqnarray}

Under an infinitesimal variation,\footnote{Note that
\bea
{\delta}\omega_{ab} &=& {\delta}\omega_{ab,c} e^c  +
\omega_{ab,c} {\delta} e^c= \bigl({\delta}\omega_{ab,c}  +
\omega_{ab,d} {\Delta}^d{}_c\bigr)e^c,\nonumber \\ 
\Rightarrow \qquad \bigl({\delta} \omega_{ab}\bigr){}_c &=& {\delta} \omega_{ab,c} 
+ \omega_{ab,d} {\Delta}^d{}_c.  \label{deltaomegai}
\eea
${\delta} \omega_{ab,c}$ is not a tensor while $\bigl({\delta} \omega_{ab}\bigr){}_c$ is.} with ${\Delta}^0{}_i=0$ maintaining the
gauge (\ref{app:Vierbeins}),
\bea
\bigl({\delta}\omega_{0i}\bigr)_0&=& -\frac{{\delta} (\partial_i  N)}{N} -\kappa_{i j} {\Delta}^j
\nonumber \\
\bigl({\delta}\omega_{0i}\bigr)_j&=& -{\delta}\kappa_{i j} - \kappa_{i k} {\Delta}^k{}_j\\
\bigl({\delta}\omega_{i j}\bigr)_0&=& -\frac{1}{N}
\bigl(\sigma_{[i j]} + \tau_{[i j]} +({\delta}\widetilde\omega_{i j})_k N^k\bigr)\nonumber \\
\bigl({\delta}\omega_{ij}\bigr)_k&=& \bigl({\delta}\widetilde \omega_{ij}\bigr)_k.
\label{deltaomega0}
\eea

Having performed the variation we can now set $\tau_{\{i j \}}=0$ for a stationary solution, but $\tau_{[i j]}$ in (\ref{omegaij}) is still arbitrary, though it must always drop out of any physical quantities. 

\subsection{Exact expression for ${\bm \Phi}[\vec K]$ \label{app:OmegaPhi}}

First we collect together all the pieces we need to calculate $\Omega$:
\begin{enumerate}
  
\item Firstly
  \bea
\int_{\partial \Sigma_t} i_{\vec {K}} \bigl({\delta} \omega_{a b} \wedge * e^{a b} \bigr)
&=& 2 \int_{\partial \Sigma_t} i_{\vec {K}} \bigl\{({\delta} \omega_a{}^b)_b *e^a\bigr\}
\nonumber \\
&=& 2 \int_{\partial \Sigma_t}
\big\{
N\bigl({\delta}\widetilde\omega_i{}^j\bigr)_j - {\delta} (\partial_i N)
- N \kappa_{i j} {\Delta}^j \nonumber \\
&& \qquad\qquad
-N_i(\kappa_{j k} \ws^{j k} + {\delta}\kappa)
\big\}* e^{0i},\label{intdeltaomwegastareoi}
\eea
where ${\delta}\kappa = {\delta}\kappa^i{}_i$ is the trace of the variation of the 
extrinsic curvature of $\Sigma_t$ in ${\cal M}$.

\item Next we need
$\int_{\partial \Sigma_t} i_{\vec {K}} d*(e^a \wedge {\delta} e_a)$.
This can be evaluated using 
\[ {\cal L}_{\vec K} {\alpha}=\partial_t {\alpha},\]
for any $p$-form $\alpha$, to write
\bea \int_{\partial \Sigma_t} i_{\vec {K}} d*(e^a \wedge {\delta} e_a)
&=&\int_{\partial \Sigma_t} {{\cal L}}_{\vec {K}}*(e^a \wedge {\delta} e_a)=
\int_{\partial \Sigma_t} \partial_t(e^a \wedge {\delta} e_a)\nonumber\\
&=&
\int_{\partial \Sigma_t} \partial_t({\Delta}_{0 i} - {\Delta}_{i 0}) * e^{0 i}
=-\int_{\partial \Sigma_t} \partial_t({\Delta}_i * e^{0 i})\nonumber\\
&=& 
-\int_{\partial \Sigma_t} \bigl( \partial_t {\Delta}_i
+ \tau^j{}_j  {\Delta}_i 
- \tau_i{}^j {\Delta}_j \bigr)* e^{0 i}\nonumber \\
&=& 
-\int_{\partial \Sigma_t} \bigl( \partial_t {\Delta}_i
- \tau_{[i j]} {\Delta}^j \bigr)* e^{0 i},\label{Kd*edeltae}\eea
where in the last step we have assumed that $\frac{\partial}{\partial t}$ is Killing so $\tau_{\{ i j\}}=0$.
Now on-shell
\[\partial_t {\Delta}_i=
    \delta \tau_{i 0}
     + \tau_{[i j]} {\Delta}^j\]
when $\frac{\partial}{\partial t}$ is Killing\footnote{Note that, although $\tau^i{}_0=0$ when $\frac{\partial}{\partial t}$ is Killing, we do not
assume that ${\delta}\tau^i{}_0=0$. We do not assume that ${\Delta}^a{}_b$ has the same symmetries as the unperturbed metric.} for $e^a$,
(so  $\dot N=0$, $\dot N^\alpha=0\Rightarrow \tau_{i 0}=0$, as well as $\tau_{\{i j\}} = 0$).
In this case (\ref{Kd*edeltae}) is simply
\beq 
\int_{\partial \Sigma_t} {{\cal L}}_{\vec {K}}*(e^a \wedge {\delta} e_a)=-\int_{\partial \Sigma_t} ({\delta} \tau_{i 0}) * e^{0 i}.
\label{dot*e_delta_e}
\eeq

\item The final piece we need to calculate $\Omega$ is ${\delta} \int * d K$.
First
\bean d K &=& \frac{1}{N}(\partial_i N^2 -2  \sigma_{i j}N^j + \dot N^i 
+ \tau_{j i} N^j )e^{0i} + \sigma_{[i j]} e^{i j}\\
\Rightarrow \qquad \int_{\partial \Sigma_t}* d K &=&
\int_{\partial \Sigma_t} \big(2\partial_i N - \frac{2}{N}  \sigma_{i j}N^j
+ \frac{1}{N}(\dot N_i +  \tau_{j i} N^j)\big)*e^{0i}\\
&=&
\int_{\partial \Sigma_t} \big(2\partial_i N +2\kappa_{i j}N^j
+ \frac{1}{N}(\dot N_i -  \tau_{i j} N^j)\big)*e^{0i}.\eean
When $\vec {K}$ is Killing 
$\tau^i{}_0=\frac{1}{N}(\dot N^i - \tau^i{}_j N^j)=0$, but in general 
\bean {\delta} (\dot N^i -  N_j \tau^{i j}  )
&=& N {\delta} \tau^i{}_0 +  ({\delta} N) \tau^i{}_0\\
&=&N{\delta} \tau^i{}_0 \qquad \hbox{on-shell}
\eean
does not vanish.

Therefore, on-shell,
\bea
{\delta}\biggl(\int_{\partial \Sigma_t}* d K\biggr)
&=& \label{delta*dK}\\
&&\kern - 100pt \int_{\partial \Sigma_t}\left\{ 2{\delta}\bigg(\partial_i N +\kappa_{i j}N^j\bigg)
+ 2\bigg(\partial_j N +\kappa_{j k} N^k\bigg)\big( \delta_i^j\ws -
{\Delta}_i{}^j\big) + {\delta} \tau_{i 0}\right\}* e^{0 i}\nonumber\\
&&\kern - 80pt
=\int_{\partial \Sigma_t}\left\{
2 {X_{i j}} N^j
+ 2 N \kappa_{ i j } {\Delta}^j
- 2 {\Delta}_i{}^j \partial_j N + 2 ( \partial_i N )  \ws
+{\delta} \tau_{i0}\right\}* e^{0 i}\nonumber
\eea
where
\beq
{X}_{i j}={\delta} \kappa_{i j} + [\kappa,{\Delta}]_{i j}
+\kappa_{i j} \ws
\eeq
with $\ws = \ws^k{}_k$ and $[\kappa,{\Delta}]_{i j}$ is the commutator of the matrices $\kappa_{i j}$ and ${\Delta}_{i j}$.

\end{enumerate}

Assembling (\ref{intdeltaomwegastareoi}), (\ref{dot*e_delta_e}) and (\ref{delta*dK}) 
one finds
\bea
\Omega[e^a,{{\cal L}}_{\vec {K}} e^a, {\delta} e^a]
&=&\frac{1}{16 \pi} \int_{\partial \Sigma_t}
\left\{2 i_{\vec {K}}\bigl({\delta} \omega_{a b} \wedge * e^{a b } \bigr)
i_{\vec {K}} d*(e^a\wedge {\delta} e_a) + {\delta} * d K \right\}\nonumber\\
&=& \frac{1}{8\pi }\int_{\partial \Sigma_t}
\big\{
N\bigl({\delta}\widetilde \omega_i{}^j\bigr)_j
-(\partial_j N)  {\Delta}_i{}^j 
+(\partial_i N)\ws\nonumber\\
&& \kern 80pt
+{X}_{i j} N^j - N_i (\kappa_{j k} \ws^{j k} + {\delta}\kappa)
\big\}* e^{0 i}.\label{Omega-1}
\eea
Now 
\[ N({\delta}\widetilde\omega_i{}^j)_j
-(\partial_j N) {\Delta}_i{}^j
= N(\wD_j \ws_i{}^j - \partial_i \ws) 
-\wD_j(N A_i{}^j)-(\partial_j N) \ws_i{}^j\]
and $\Omega[e^a,{{\cal L}}_{\vec {K}} e^a, {\delta} e^a]$ should be gauge invariant and hence independent of
$ A_i{}^j$.
We might expect
\beq\int_{\partial \Sigma_t} \wD_j (N A_i{}^j)  * e^{0 i}=0\label{Frobenius}\eeq
as it is the integral of a divergence on $\partial \Sigma_t$
(the integral over $*e^{0 i}$ forces $e^i$ to be normal to $\partial \Sigma_t$ 
and hence $j$ is restricted to be a tangential index).  
This can be proven more rigorously using a straightforward application of Frobenius' theorem but we omit the details.
Invoking (\ref{Frobenius}) equation (\ref{Omega-1}) can be re-arranged to
\bea
\Omega[e^a,{{\cal L}}_{\vec {K}} e^a, {\delta} e^a]
&=& \frac{1}{8\pi }\int_{\partial \Sigma_t}
\big\{
N\bigl(\wD_j \ws_i{}^j - \partial_i \ws)
+(\partial_i N) \ws - (\partial_j N)\ws_i{}^j\nonumber\\
&&\kern 50pt
+{X}_{i j} N^j - N_i (\kappa_{j k} \ws^{j k} + {\delta}\kappa)
\big\}* e^{0 i}.\label{Omega-2}
\eea

Assuming that $\Sigma_t$ can be foliated into 2-dimensional surfaces $\phi_{t,r}$, parameterised by $r$, we have 
\bea
\Phi[e^a,{{\cal L}}_{\vec {K}} e^a, {\delta} e^a]
&=& \frac{1}{8\pi }\int_{\sigma_{t,r}}
\big\{
N\bigl(\wD_j\ws_i{}^j - \partial_i \ws)
+(\partial_i N) \ws - (\partial_j N)\ws_i{}^j\nonumber\\
&&\kern 50 pt
+{X}_{i j} N^j - N_i (\kappa_{j k} \ws^{j k} + {\delta}\kappa)
\big\}* e^{0 i}.\label{Omega_partial_Sigma_b}
\eea
is guaranteed to be independent of $t$ and $r$ if $\vec {K} = \frac{\partial}{\partial t}$ is Killing.

\section{Connection 1-forms in Bondi-Sachs gauge\label{app:B-S}}

We list here the connection 1-forms for asymptotically flat metrics in Bondi-Sachs the gauge as
defined in (\ref{BS-gauge1}) and (\ref{C-matrix}) in the text.
First write
\[ f= U \exp W \qquad \hbox{and} \qquad g=U^{-1} \exp(W)\]
with $f(u,r,\vartheta,\varphi)$  and $g(u,r,\vartheta,\varphi)$ tending to one as $r\rightarrow\infty$
in terms of which
\beq e^0=f du + g d r,\qquad e^1=g dr.\label{Be0e1}\eeq
We also have
\beq e^i = r\tilde e^i - V^i du\label{Be2e3} \eeq
with  $\tilde e^{i}$ ($i=2,3$) orthonormal 1-forms for the 2-sphere metric with $r$ and $u$ constant.
In terms of the round unit metric on the 2-sphere it is convenient to choose
a gauge in which
\[ \tilde e^i = C^i{}_j \hat e^i\]
with
\[ C^i{}_j = \begin{pmatrix}
e^\gamma \cosh \delta & e^{-\gamma} \sinh \delta \\
e^\gamma  \sinh \delta & e^{-\gamma} \cosh \delta
\end{pmatrix}\]
and \[ \hat e^2 = d\vartheta, \qquad \hat e^3 =\sin\vartheta d\varphi.\]
$\det C=1$, but we shall not need this explicit form.

The connection 1-forms arising from (\ref{Be0e1}) and (\ref{Be2e3}) can be calculated from (\ref{omega-ide}),
they are\footnote{$\{ i j \}$ denotes symmetrisation, with normalisation  $\frac{1}{2}(i j +  j i)$; $[i j]$ denotes anti-symmetrisation, with normalisation $\frac{1}{2}(i j -  j i)$}

\bean
\omega_{0 1}&=&\frac{1}{f g} \big( \dot g +  V^i \partial_i g\big) (e^0-e^1) - \frac{f'}{f g} e^0
-\frac{1}{2} \left( \frac{\partial_i f}{f}-  \frac{\partial_i g}{g} +\frac{V_i'}{f g} \right) e^i ,\\
\omega_{0 i}&=&-\frac{\partial_i f}{f}e^0 
+\frac{1}{2}\left(\frac{\partial_i f}{f} - \frac{\partial_i g}{g}  - \frac{V_i'}{f g}\right)e^1 
 -\frac{1}{f} \big(\widetilde D_{\{i} V_{j\}} + \tau_{\{i j\} }  \big) e^j ,\\
\omega_{1 i}&=&
 \frac{\partial_i g}{g}e^1+\frac{1}{2}\left(\frac{\partial_i f}{f} - \frac{\partial_i g}{g}  + \frac{V_i'}{f g}\right)e^0 
 +\left( \frac{1}{f} \big(\widetilde D_{\{i} V_{j\}} + \tau_{\{i j\} }  \big)
-\frac{1}{g} \Big(\frac{\delta_{i j}}{r} +\rho_{\{i j  \} }\Big)\right) e^j ,\\
\omega_{i j}&=& \frac{1}{r}\widetilde \omega_{i j}+
 \frac{1}{f} \big(\widetilde D_{[i} V_{j]} - \tau_{[i j] }  \big)(e^0-e^1)
-\frac{1}{g} \rho_{[i j] }  e^1 ,
\eean
where $\dot{\ } = \partial_u$, ${\ } '=\partial_r$, $\partial_i = (\tilde e{}^{-1})^\alpha{}_i \partial_\alpha$
and
\[ \rho^i{}_j = (C' C^{-1})^i{}_j.\] 
$\widetilde \omega_{i j}=\widetilde \omega_{i j,k}\tilde e^k$ are the connection 1-forms associated with $\tilde e^i$ on the 2-sphere
and $\widetilde D_i$ the associated co-variant derivative.

With $f= U e^W$ and $g=U^{-1} e^W$ these expressions are used to calculate $\Phi$ in \S\ref{sec:BondiMass}.

Note that we have not assumed any symmetries, in particular it has not been assumed that $\vec K=\partial_t$ is Killing.  When $\partial_t$ is Killing $\dot g$ and $\tau_{\{ i j\}}$ are zero, but in any case these do not appear in $\Phi(\vec K)$ at large $r$.

\end{document}